\documentclass[conference]{IEEEtran}
\IEEEoverridecommandlockouts
\pdfoutput=1
\usepackage[english]{babel}
\usepackage{amsthm}
\newtheorem{Rule}{Rule}
\newtheorem{theorem}{Theorem}[section]

\usepackage{cite}
\usepackage{amsmath,amssymb,amsfonts}
\usepackage{algorithm}
\usepackage[noend]{algorithmic}
\usepackage{graphicx}
\usepackage{textcomp}
\usepackage{xcolor}
\usepackage{subfigure}

\usepackage{float}

\usepackage[a4paper, total={184mm,239mm}]{geometry}
\def\BibTeX{{\rm B\kern-.05em{\sc i\kern-.025em b}\kern-.08em
    T\kern-.1667em\lower.7ex\hbox{E}\kern-.125emX}}
\begin{document}

\title{DeFT: A \underline{De}adlock-Free and \underline{F}ault-\underline{T}olerant\\ Routing Algorithm for 2.5D Chiplet Networks \vspace{-0.1in}}

\author{\IEEEauthorblockN{Ebadollah Taheri, Sudeep Pasricha, and Mahdi Nikdast}
\IEEEauthorblockA{Department of Electrical and Computer Engineering, Colorado State University, USA}\vspace{-1in}}

% \and
% \and
% \and
% \and
% \and

\maketitle

% \makeatletter
% \def\normalsize{\@setfontsize{\normalsize}{9.5bp}{12.00pt}}
% \normalsize
% \makeatother

\begin{abstract}
By interconnecting smaller chiplets through an interposer, 2.5D integration offers a cost-effective and high-yield solution to implement large-scale modular systems. Nevertheless, the underlying network is prone to deadlock, despite deadlock-free chiplets, and to different faults on the vertical links used for connecting the chiplets to the interposer. Unfortunately, existing fault-tolerant routing techniques proposed for 2D and 3D on-chip networks cannot be applied to chiplet networks. To address these problems, this paper presents the first deadlock-free and fault-tolerant routing algorithm, called \textit{DeFT}, for 2.5D integrated chiplet systems. \textit{DeFT} improves the redundancy in vertical-link selection to tolerate faults in vertical links while considering network congestion. Moreover, \textit{DeFT} can tolerate different vertical-link-fault scenarios while accounting for vertical-link utilization. Compared to the state-of-the-art routing algorithms in 2.5D chiplet systems, our simulation results show that \textit{DeFT} improves network reachability by up to 75\% with a fault rate of up to 25\% and reduces the network latency by up to 40\% for multi-application execution scenarios with less than 2\% area overhead. 
\end{abstract}\vspace{-0.2in}

%\begin{IEEEkeywords}
%2.5D Chiplet Platforms, NoCs, Routing Algorithms, Deadlock Freedom, Fault-Tolerance, Energy-Efficiency.
%\end{IEEEkeywords}

\section{Introduction}
% •	Scalability and yield of 2D and 3D NoCs
% •	Why 2.5 D NoC and its routing algorithms
The continuous demand for higher computation power necessitates further scalability improvements in systems-on-chip (SoCs). Unfortunately, conventional 2D SoCs suffer from low scalability because of their low yield, and hence a high manufacturing cost when scaling up to support higher complexities \cite{kannan2015enabling}. To this end, 3D integration partitions an SoC into multiple smaller dies that can be vertically stacked using through-silicon vias (TSVs). However, 3D integration can result in a high fabrication cost and power density, and low cooling conductivity, creating thermal hotpots and degrading the reliability \cite{taheri2019addressing}. To alleviate these issues, the 2.5D integrated approach presents a modular solution by placing several chiplets on an interposer \cite{kim2019architecture}, on which inter-chiplet communication is supported. In addition, and similar to 3D SoCs, in such a modular integration, each chiplet can be designed heterogeneously \cite{ma2021tap} and independently in a short design-time, as off-the-shelf chiplets can be integrated on an interposer \cite{kim2019architecture}. \par

% •	Conventional routing algorithm for deadlock freedom
However, 2.5D integration introduces two main challenges, including deadlock avoidance and low reliability due to vertical-link (VL) faults. First, the underlying network in a 2.5D chiplet system (i.e., the intra- and inter-chiplet interconnect) suffers from deadlock, when a cyclic dependency of requests occurs in the network. Conventionally, to avoid deadlock, some turns can be restricted to break the cyclic dependency in a network \cite{ebrahimi2017ebda}. However, even when deadlock-freedom is guaranteed in intra-chiplet networks using conventional approaches \cite{ebrahimi2017ebda}, the inter-chiplet packets can still create some dependencies, and hence deadlock. Second, enabling link and router fault-tolerance is essential in 2.5D chiplet systems \cite{wang2016pre}, but it has not been addressed. Existing fault-tolerant solutions in 2D and 3D networks cannot be applied to 2.5D networks, where deadlock-freedom is more challenging due to higher irregularity in the network. In particular, enabling fault-tolerance in 2.5D chiplet systems requires higher path redundancy to be supported in the routing algorithm, which makes the deadlock-freedom and load-distribution even more complex. Current routing algorithms~\cite{yin2018modular,majumder2020remote} to address the deadlock in 2.5D chiplet systems limit the VL selection, and hence deteriorate network reliability and performance.  \par

%Current routing algorithms  offer a low adaptation in VL selection and, as a result, suffer from low reliability.  

%In \cite{yin2018modular}, a modular-turn restriction (MTR) routing algorithm was proposed where the routing on each chiplet is designed separately by avoiding some turns from the chiplet to the interposer and vice versa. However, interposer is supposed to be informed of the restricted turns of chiplets. This violates a key attribute in 2.5D systems to enable the design of each chiplet and interposer independently without the knowledge of the whole design. \cite{majumder2020remote}. To this end, in Remote Control (RC) routing algorithm \cite{majumder2020remote}, deadlock freedom of chiplet systems is guaranteed by reserving an extra buffer on the chiplet routers connected to interposer, a.k.a. boundary routers. Both routing algorithms limit vertical link (VL) selection, which deteriorates network reliability and traffic distribution.
   
% •	Paper contribution
The novel contribution of this paper is on developing the first \underline{De}adlock-free and \underline{F}ault-\underline{T}olerant routing algorithm, called \textit{DeFT}, for 2.5D chiplet systems. In particular, deadlock-freedom is guaranteed by employing a novel virtual-network (VN) assignment strategy to ensure that the network virtual-channel (VC) utilization is highly balanced. In addition, \textit{DeFT} proposes a novel dynamic VL-selection strategy that improves both the fault-tolerance and the load-distribution on the VLs to reduce the network latency. Our results show that \textit{DeFT} can achieve 100\% reachability under different VL-fault scenarios (e.g., 25\% faults) in chiplet systems, while related work can only tolerate less than 3.1\% and 2.1\% faults on VLs for systems with 4 and 6 chiplets, respectively. Moreover, \textit{DeFT} improves the latency under real-application traffic by 3\% and 13.5\%, on average, for relatively low and high traffic scenarios, respectively, with less than 2\% area overhead.   \par 
%unless chiplet is completely disconnected from interposer i.e no fault-free VL remains, 

% •	Paper organization
%The rest of the paper is organized as follows. We review some related work and the motivation behind this work in Section II. Section III presents the proposed \textit{DeFT} routing algorithm and its novel solutions for deadlock-freedom, fault-tolerance, and congestion-aware VL selection. We discuss our evaluation and simulation results in Section IV. Finally, Section V concludes the paper. \par
\begin{figure}[t]
\centering
\includegraphics[scale=0.34]{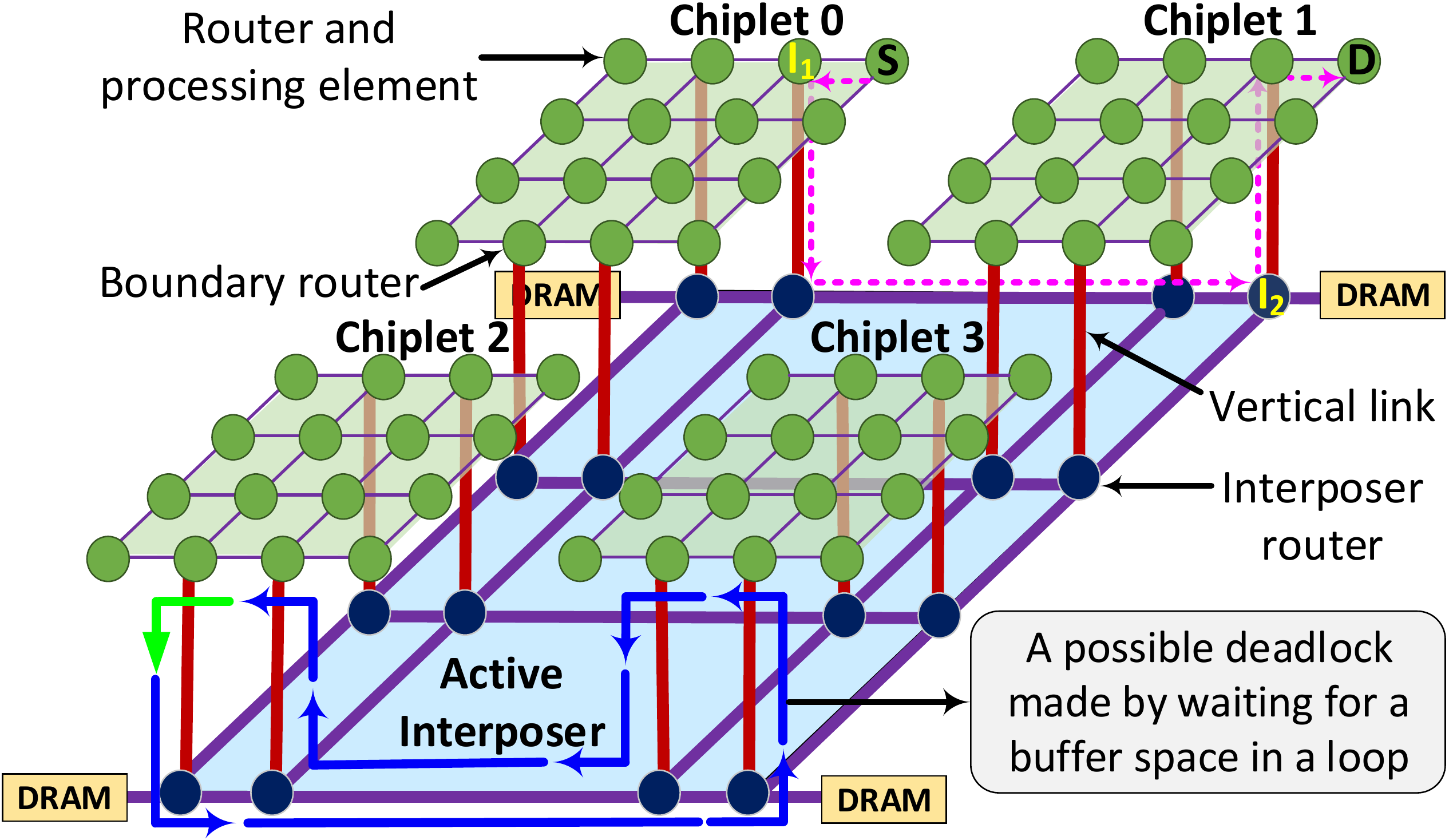}
\vspace{-0.15in}
\caption{\small An abstract overview of the baseline 2.5D network with four chiplets on an active interposer.}
\vspace{-0.2in}
\label{Bseline}
\end{figure}

\section{Background and Related Work}
\subsection{Deadlock-Free Routing in 2.5D Chiplet Systems}
Fig.~\ref{Bseline} shows the baseline 2.5D chiplet system considered in this paper with 4 CPU chiplets on an active interposer. Many integration approaches exist in chiplet systems \cite{kannan2015enabling}, including different combinations of CPU, GPU, and DRAM chiplets. We consider Fig.~\ref{Bseline} as our case study, but our approach can be employed in any chiplet system. In Fig.~\ref{Bseline}, each chiplet is connected by 4 bidirectional VLs to the interposer, where the routers on chiplets attached to the VLs are called \textit{boundary routers}. Routing in chiplet-based systems uses two intermediate destinations: one on the source chiplet and the other one on the interposer. For example, in Fig.~\ref{Bseline}, the pink routing path from S on Chiplet 0 to D on Chiplet 1 uses I$_1$ and I$_2$ as intermediate destinations. On the source chiplet, a VL is selected as the first intermediate destination. Then, the packet is routed to the selected VL, and from there it is vertically routed to the interposer. Similarly, the second intermediate destination is selected on the interposer for the packet to get to its destination chiplet. Finally, the packet can be routed to its eventual destination on the destination chiplet (Fig. 1). \par

%Assuming wormhole switching, an incoming packet enters a router's buffer and moves forward by selecting an output port based on the routing algorithm and the next router's buffer availability, where waiting for the availability can lead to a cyclic dependency (i.e., deadlock). 
Deadlock occurs when packets are in a cyclic dependency, waiting on each other to release reserved buffers. For instance, dimension-order routing algorithms(e.g. XY routing in 2D mesh) are deadlock-free \cite{glass1992turn}, because the allowed turns in the network cannot make a cycle. While intra-chiplet routing can be deadlock-free (e.g., using XY routing or conventional turn models \cite{ebrahimi2017ebda}), cyclic dependencies can occur when integrating several of these chiplets on an interposer, leading to deadlock. Fig.~\ref{Bseline} shows an example for a deadlock scenario occurring from the combination of some turns (shown with blue and green arrows) between deadlock-free Chiplets 2 and 3. \par

% •	Modular routing and its drawbacks (main: vertical selection adaptation)
Several routing algorithms have been recently proposed to overcome deadlock in mesh-based 2.5D chiplet systems, namely the modular-turn restriction (MTR) \cite{yin2018modular} and the remote control (RC) \cite{majumder2020remote} algorithms. MTR employs turn restrictions to avoid some inter-chiplet turns, on the boundary routers, and break cyclic dependencies. As an example in Fig.~\ref{Bseline}, avoiding the left-to-down turn in Chiplet 2 (shown with the green arrow) can break the cyclic dependency. However, MTR makes the interposer router and chiplet designs dependent, violating the modular design requirement in chiplet-based systems, because each interposer router needs to know whether a packet can reach its destination through a VL while considering the restricted turns. To this end, RC routing \cite{majumder2020remote} breaks the inter-chiplet cyclic dependency by employing an extra buffer (RC-buffer) on the boundary routers to store the whole packet. The RC-buffer is shared among the chiplet routers that utilize the boundary router for inter-chiplet communication. However, RC requires extra hardware for the RC-buffer and a permission network as the RC-buffer is shared among several routers. In addition, both MTR and RC limit VL selection, which restricts them to re-select VLs and makes them unable to tolerate VL faults. Other works on chiplet systems, such as \cite{kannan2015enabling, bharadwaj2020kite}, only briefly discuss routing algorithms and consider using virtual-channel-based deadlock avoidance with unbalanced utilization of virtual channels while ignoring VL faults.

% •	Fault tolerant NoCs 
% •	Conventional fault tolerant routing algorithms, 2D and 3D networks, are not useful in 2.5D
\subsection{Fault-Tolerant Routing in 2.5D Chiplet Systems}
To the best of our knowledge, there is no prior work on addressing VL faults in 2.5D chiplet systems. In addition, existing routing algorithms proposed for 3D networks---for both fully \cite{pasricha2011low, ebrahimi2013fault} and partially \cite{salamat2016resilient, taheri2019addressing} connected networks---cannot be applied to 2.5D chiplet systems because: 1) in 3D networks, a packet goes from one direction to another vertically (up$\rightarrow$down or down$\rightarrow$up), while in a 2.5D network, packets go down and then up (up$\rightarrow$down$\rightarrow$up); i.e., an inter-chiplet packet requires two vertical routings and three intra-layer routings: on the source chiplet, interposer, and destination chiplet; 2) some routers are indirectly connected to other chiplets, and the chiplet and interposer sizes may also be different, which makes the topology more irregular than 3D networks. \par

% •	Highlight paper contributions
Unlike conventional 2.5D routing
algorithms \cite{yin2018modular, majumder2020remote} \textit{DeFT}, does not put any limitation on VL selection to realize deadlock-freedom. Therefore, any VL-fault patterns can be tolerated unless the network connectivity is not guaranteed (i.e., chiplets are disconnected). Moreover, \textit{DeFT} includes a novel congestion-aware VL-selection strategy to improve traffic distribution especially in the presence of VL faults.

\begin{figure}[t]
\centering
\includegraphics[scale=0.25]{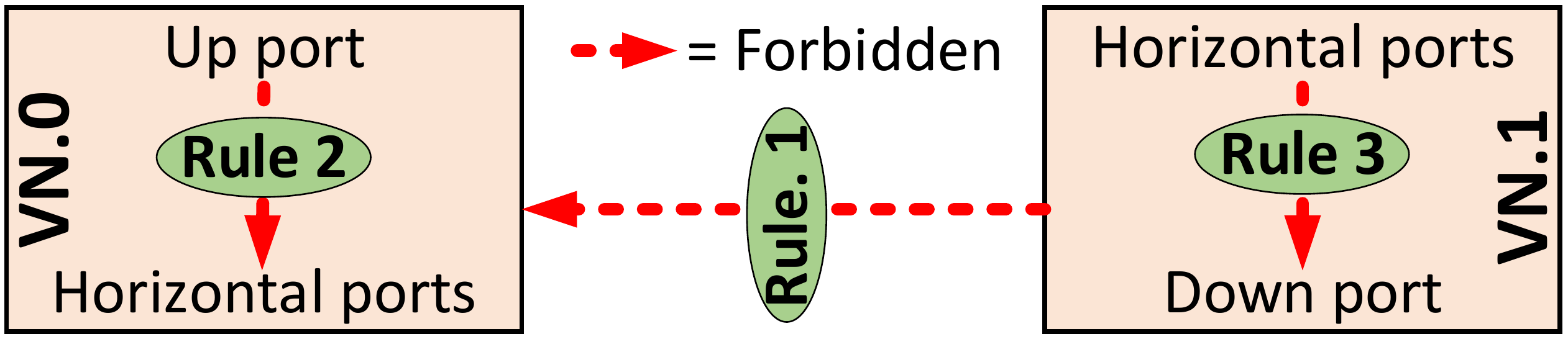}
\vspace{-0.1in}
\caption{\small \textit{DeFT}'s rules for VN utilization and deadlock-freedom.}
\vspace{-0.25in}
\label{Rules}
\end{figure}

\section{\textit{DeFT}: Proposed Routing Algorithm}
%This section discusses our VN-separation approach to enable deadlock-freedom, and our fault-tolerant VL-selection strategy. 

% •	Motivation and challenge in deadlock freedom
% •	Deadlock freedom with virtual channel
% •	The algorithm
\subsection{Proposed VN Separation and Deadlock-Freedom}
The idea of employing VN separation for deadlock-freedom is relatively old. However, employing VNs while considering hardware overhead and network latency is quite challenging. Consequently, existing routing techniques\cite{yin2018modular, majumder2020remote} in chiplet systems have underestimated the efficiency of VN separation for deadlock-freedom. As we will show, \textit{DeFT} employs two VCs with negligible hardware overhead and fair utilization of VCs, to realize deadlock-freedom and improve hardware efficiency and traffic distribution.

 %as DeFT employ two VCs to have a  utilization-aware deadlock freedom.

\textit{DeFT} utilizes two VNs for deadlock-freedom, where at least one VC is required for each VN. Here, we assume one VC per VN, but the number of VCs can be increased without loss of generality. We define ``Down" port as the one going from a chiplet to the interposer, ``Up" port goes from the interposer to a chiplet, and ``Horizontal" ports (East, West, South, and North) are intra-chiplet and intra-interposer ports. The following rules, also shown in Fig.~\ref{Rules}, are used for deadlock-free VN utilization:

\begin{Rule}
Routing from VN.1 to VN.0 is forbidden, while packets can go from VN.0 to VN.1.\vspace{-0.05in}
\end{Rule}
\begin{Rule}
For packets in VN.0, routing from an Up port to Horizontal ports is forbidden.\vspace{-0.05in}
\end{Rule}
\begin{Rule}
For packets in VN.1, routing from Horizontal ports to a Down port is forbidden.\vspace{-0.05in} 
\end{Rule}

To satisfy these three rules, for \emph{inter-chiplet packets injected from the non-boundary routers}, VN.0 is assigned to the packets in the source router. Such VN.0 assignment should be changed to VN.1 (VN0$\rightarrow$VN1) between the first (leaving the source chiplet) and the second (entering the destination chiplet) vertical routing. While these rules guarantee deadlock freedom, VN utilization is efficient due to the following theorems:
\vspace{-0.05in}
\begin{theorem}
\label{theorem1}
Both VNs can be assigned to intra-chiplet packets.\vspace{-0.05in}
\end{theorem}

\begin{proof}
Intra-chiplet packets do not use vertical ports (Up or Down ports), and hence they do not face any forbidden routings in VNs stated in Rules 2 and 3. \vspace{-0.05in}
\end{proof}

\begin{theorem}
\label{theorem2}
Both VNs can be assigned to inter-chiplet packets for routing on the interposer.
\end{theorem}
\begin{proof}
Packets entering the interposer are in VN.0. They can remain in VN.0 based on Rule 2 (Horizontal to Down is not forbidden in VN.0) or, based on Rule 1, switch to VN.1.\vspace{-0.05in}
\end{proof}
%Packets entering the interposer are in VN.0. Going from Horizontal of source chiplet to Down, they can remain in VN.0 based on Rule 2 (Horizontal to Down is not forbidden in VN.0) or, based on Rule 1, switch to VN.1. Going from Down to Horizontal of the interposer, based on Rules 2 and 3, they do not face and forbidden routing
\begin{theorem}
\label{theorem3}
Inter-chiplet packets on the source chiplet are free to be routed via any VLs towards the interposer.\vspace{-0.05in}
\end{theorem}
\begin{proof}
Packets from a source chiplet to the interposer need to go from the Horizontal ports to the Down port of the selected VL, and then from the Down port to the Horizontal ports of the interposer. Inter-chiplet packets on the source chiplet are in VN.0. When using any VL to go to the interposer, 1) if the packet stays in VN.0, based on Rule 2, routing from Horizontal to Down ports, and Down to Horizontal ports are not forbidden; and 2) if the packet is switched to VN.1, based on Rule 1, it can to go to a Down port and, based on Rule 3, it can go from Down to Horizontal ports.\vspace{-0.05in}   
\end{proof}

\begin{theorem}
\label{theorem4}
Inter-chiplet packets on the interposer are free to be routed via any VLs towards the destination chiplet.\vspace{-0.05in}
\end{theorem}
\begin{proof}
Routing from the interposer to the destination chiplet is done by going from Horizontal ports to the Up port of the selected VL, and then from the Up port to Horizontal ports of the destination chiplet. Based on Rules 2 and 3, regardless of their VN, packets can go from Horizontal to the Up port. When using any VL to go to the destination chiplet, 1) if a packet is in VN.0, based on Rule 1, it can be switched to VN.1 to go from Up to Horizontal ports; and 2) if the packet is in VN.1, based on Rule 3, it can go from Up to Horizontal ports.\vspace{-0.05in}
\end{proof}

% •	Vertical selection adaptation
Based on Theorems \ref{theorem1} and \ref{theorem2}, the proposed VN separation offers a high balance of VC utilization to improve traffic distribution on VCs while also utilizing a small number of VCs (i.e., two VCs in total). For the cases where both VNs can be assigned, we use round-robin assignment to balance the VN load. Moreover, according to Theorems \ref{theorem3} and \ref{theorem4}, the choices for VL selection are maximized to tolerate faults on VLs. In addition, as we will discuss in Section \ref{SelectionSection}, such flexibility in VL selection helps balance load-traffic on VLs and improves the network latency. Algorithm~\ref{Algorithm_VN} summarizes our VN-assignment strategy. Note that, as shown in Algorithm~\ref{Algorithm_VN} and Fig.~\ref{Bseline}, an interposer router can also be a source router (e.g., for the packets injected by DRAMs).\vspace{-0.05in} 
\begin{algorithm}
\caption{Virtual-Network (VN) Assignment in \textit{DeFT}}\label{Algorithm_VN}
\begin{algorithmic}
\small
%\IF{current is source \textbf{\&} destination is on different chiplet}
%\ELSE[$N$ is odd]
\IF{Current router is Source}
    \IF{Source $\in$ \{interposer $\cup$ dest. chip $\cup$ boundaries\}}
    \STATE Do round-robin assignment between VN.0 and VN.1
    \ELSIF{Destination is on a different chiplet}
    \STATE Assign VN.0
    \ENDIF
%\ENDIF
\ELSIF{Current router $\in$ boundary routers}
\IF{going to the interposer}
\STATE Do round-robin reassignment between VN.0 and VN.1
\ELSIF{coming from the interposer}
\STATE Go to (remain in) VN.1
\ENDIF
\ELSE
\STATE Stay in the previously assigned VN 
\ENDIF
%\STATE $C_e \s $ Calculate 
\end{algorithmic}
\end{algorithm}\vspace{-0.05in} 

%\subsection{Deadlock and livelock freedom}
% •	Deadlock definition and freedom proof of proposed routing
\textbf{Deadlock-freedom:} \textit{DeFT} is deadlock-free in any 2.5D chiplet system where each chiplet is locally deadlock-free, because of two main reasons: 1) there is no inter-chiplet cyclic dependency in VN.0 and VN.1. In VN.0, based on Rule 2, routing from Up port to Horizontal ports is avoided. In VN.1, based on Rule 3, routing from Horizontal ports to Down port is avoided; and 2) there is no cyclic dependency between VNs because Rule 1 avoids routing from VN.1 to VN.0. %Considering these reasons, \textit{DeFT} is deadlock-free because no cycle can occur.      

% •	Livelock definition and freedom proof of proposed routing
\textbf{Livelock-freedom:} \textit{DeFT} is livelock-free in any 2.5D chiplet system where each chiplet is locally livelock-free.
The only possible routing for inter-chiplet packets is source-chiplet $\rightarrow$ interposer $\rightarrow$ destination-chiplet. This is done by the two intermediate destinations (see Section II-A) where the packets are minimally routed between the intermediate and the main destinations. Intra-chiplet packets are also routed minimally. Therefore, packets are routed in a finite number of hops, and hence \textit{DeFT} is livelock-free.\vspace{-0.05in}

\begin{figure}[t]
\centering
\includegraphics[scale=0.5 ]{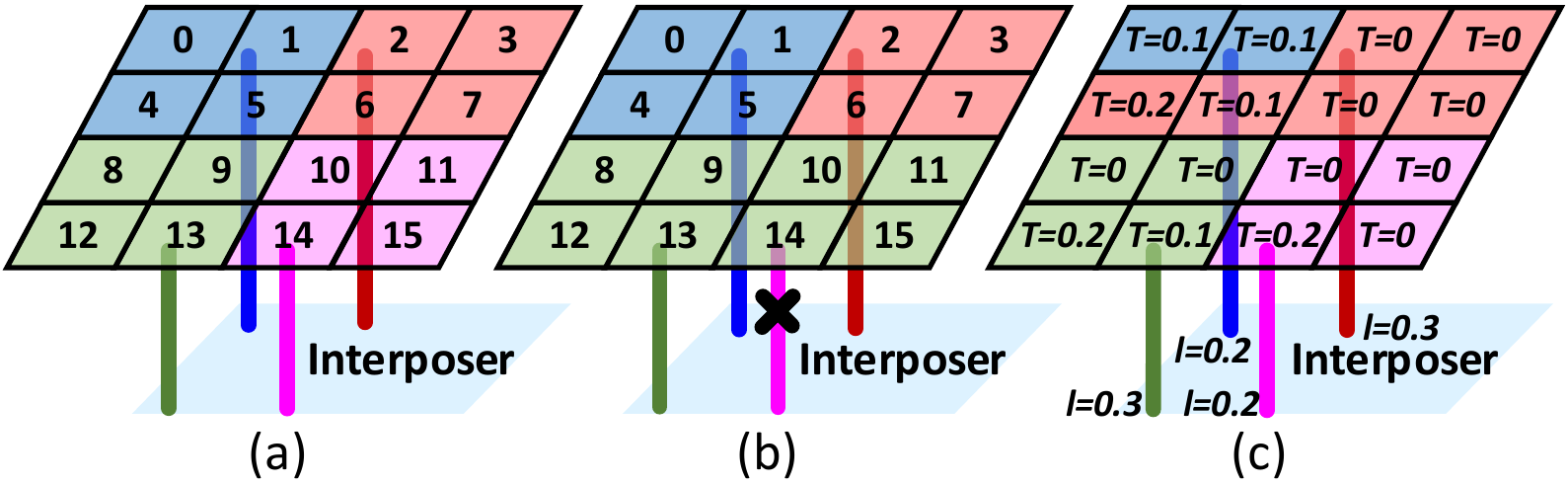}
\vspace{-0.15in}
\caption{\small Examples for a VL selection: (a) a fault-free distance-based selection (closest VL), (b) a distance-based selection in the presence of a VL fault, and (c) a good selection under non-uniform traffic. $T$ and $l$ denote the inter-chiplet traffic rate of routers and VLs, respectively. Each tile is a router and router's color represents its selected VL.}
\label{SelMotivation}
\vspace{-0.25in}
\end{figure}

\subsection{Proposed Fault-Tolerant Congestion-Aware VL Selection}
\label{SelectionSection}
%\subsection{Motivation to an optimized fault tolerant selection}
% •	Motivation for traffic- and energy-aware selection
% •	Advantage of vertical link selection adaptation in the high-performance selection
% •	The algorithm
%As we already discussed, for inter-chiplet packets, two temporary destinations are selected through VL selection processes. Based on a selection algorithm, a VL is selected and the packet is forwarded to the VL to get to the other chip. 

Considering VL faults---due to inevitable mismatch \cite{li20173d}, electromigration \cite{hsiao2015electromigration}, and thermomigration \cite{li20173d} in microbumps---is essential in 2.5D chiplet systems. For the first time, \textit{DeFT} takes on this challenge that has been ignored in existing routing algorithms. To tolerate VL faults, affecting inter-chiplet routing, \textit{DeFT} supports an adaptive VL selection where VLs are adaptively selected as intermediate destinations (see Section II-A). Fig.~\ref{SelMotivation} shows several examples of VL selection in a chiplet with four VLs where the color of each router (shown as tiles) indicates its selected VL. Here, $T$ and $l$ denote the inter-chiplet traffic rate of routers and VLs, respectively. Traffic rate of a VL ($l$) is the sum of the traffic rates of the routers ($T$ of routers) selecting the VL. 

Fig.~\ref{SelMotivation}(a) shows an example of a fault-free distance-based selection (i.e., selecting the closest VL for each router) where the distribution of VLs and traffic pattern are both uniform. In Fig.~\ref{SelMotivation}(b), in which one VL is faulty, a distance-based selection approach decides that, for example, eight routers should utilize the green VL while the two other VLs are utilized by four routers each. But such an unbalanced utilization can significantly increase the chance of congestion, degrading the overall performance. A more efficient selection in this example would be for Routers 8 and 11 to utilize the blue and the red VL, respectively. That way, for example, although the blue VL is farther to Router 8, packets injected by Router 8 can benefit from a lower traffic load on the blue VL and latency. 

In addition to VL faults, an efficient VL selection should consider the traffic profile into account to properly distribute the load on VLs \cite{Adele}. Fig.~\ref{SelMotivation}(c) shows an example where an efficient selection is done based on the traffic profile. Applying distance-based selection (similar to Fig.~\ref{SelMotivation}(a)) to Fig.~\ref{SelMotivation}(c), the loads on each VL will be $l$\textsubscript{blue}$=$0.5, $l$\textsubscript{red}$=$0, $l$\textsubscript{green}$=$0.3, and $l$\textsubscript{purple}=$0.2$. This puts half of the total load on the blue VL while there is no load on the red VL. \textit{DeFT} considers both the VL faults and traffic profile to enable a fault-tolerant and congestion-aware VL selection, as discussed next.
%(b) VL faults change healthy VL distribution where even with uniform traffic distance based selection put high load on green VL, (c)

The VL-selection strategy in \textit{DeFT} includes an offline step to analyze optimal VLs under different VL-fault scenarios and an online selection among such pre-analyzed VLs when a fault occurs. During design-time and considering VL-utilization balancing and distance (source to VL hop-count) minimization, we explore different VL selections for all the possible VL-fault scenarios. VL-utilization balancing is important because it relieves over-utilized VLs from extra load. It not only improves the network latency but also increases the system reliability, as over-utilization of VLs can increase stress-migration-based faults \cite{hsiao2015electromigration}. Moreover, distance minimization can also improve energy consumption, as reducing the path length of a packet can reduce dynamic power dissipation in routers due to that packet, as well as its end-to-end latency. Considering these two objective functions, the best resulting VL selections are analyzed and saved in look-up tables in routers to be utilized during run-time. This improves hardware complexity when calculating an optimized selection, providing \textit{DeFT} with an adaptive selection under different VL-fault scenarios. Here, the look-up table size and hardware cost is negligible compared to the overall size of a router (see Section~\ref{hardwareSec} for cost analysis).

For any inter-chiplet packet, two VL selections are required: one on the source chiplet (towards the interposer) and one on the interposer (towards the destination chiplet). The first selection is done based on the packets generated from the source-chiplet routers. For VL selection on the interposer, packets headed for the destination-chiplet routers affect the selection. As both VL selections are similar, here we only discuss VL selection on the source chiplet for brevity. Given a fault scenario and based on the VL selection on the source chiplet, i.e., the first selection, we consider the source-chiplet routers ($r$) which are included in the VL-selection process: $R_C=\{r_1, r_2, .., r_R\}$. Selection is done per chiplet, where $C$ is the chiplet and $R_C$ is the chiplet's routers. The chiplet is connected to the interposer using a set of fault-free vertical links ($v$): ${VL}_C=\{v_1, v_2,..,v_V\}$. The objective is to find a set of optimized VL selections for all the source-chiplet routers while balancing the VL utilization: $s^*=\{V_{r_1},V_{r_1},...,V_{r_R}\}$, where $V_{r_i}$ is the VL selected for router $i$.

To balance VL utilization, the load on each VL should be close to the average load on all the VLs. The load on VL$_i$ depends on the inter-chiplet traffic (packet injection) rate among the routers that select VL$_i$. The load on VL $v$ is: \vspace{-0.05in}
\begin{equation}
l_v = \sum_{r\in R_C} T^{inter}_r \times U^v_r,\label{VL-load}\vspace{-0.08in}
\end{equation}
where $T^{inter}_r$ is the inter-chiplet traffic rate of router $r$. Also, $U^v_r=1$ ($U^v_r=0$) whenever router $r$ utilizes (does not utilize) VL $v$ for vertical routing. Based on \eqref{VL-load}, the average load over all the VLs is:\vspace{-0.05in}
\begin{equation}
    l_{avg} = \frac{1}{V}\sum_{v\in {VL}_C} l_v.\label{avg-load}\vspace{-0.05in}
\end{equation}
Considering \eqref{VL-load} and \eqref{avg-load}, we define the load cost of a VL $v$ as:\vspace{-0.05in}
\begin{equation}
L_{v} = \left |\frac{l_v - l_{avg}}{l_{avg}}\right|. \label{LoadCost_Eq}\vspace{-0.05in}
\end{equation}

The second objective in VL selection is distance minimization. Considering a mesh network, the distance (i.e., hop count) between a router $r$ to a VL $v$ (on the same chiplet) is:\vspace{-0.05in}
\begin{equation}
    D^{r}_{v} = \left | x_r-x_v \right |+ \left | y_r-y_v\right|,\label{r-v-distance}\vspace{-0.05in}
\end{equation}
where $x_r$ and $y_r$ are the coordinates of the router $r$, and $x_v$ and $y_v$ are the coordinates of the VL $v$.
The distance cost of a VL $v$ to the routers that select $v$ is:\vspace{-0.05in}
\begin{equation}
D_v = \sum_{r\in R_C} D^{r}_{v} \times U^v_r.\label{DisCost_Eq}\vspace{-0.05in}
\end{equation}
Based on \eqref{LoadCost_Eq} and \eqref{DisCost_Eq}, the overall cost of a selection set $s$ is:\vspace{-0.05in}
\begin{equation}
C_s = \sum_{v\in {VL}_C} (\rho \times D_{v}) + L_{v},\label{OverallCost_Eq}\vspace{-0.05in}
\end{equation}
where $\rho$ can decide the importance of the load-balancing versus distance objectives. In our analyses (see Section~\ref{results}), we experimentally found $\rho=$~0.01 to be efficient in \textit{DeFT}. Based on the cost function $C_s$ in \eqref{OverallCost_Eq}, an optimization search $O$ can be used to find the optimal selection set $s^*$ with the minimum cost $C_s^*$ (Algorithm~\ref{Algorithm_sel}):\vspace{-0.05in}
\begin{equation}
s^*\gets O(s\in S,\textrm{Objective}:\min(C_s)).\label{Optimization_Eq}\vspace{-0.05in}
%s^*=\mathop{\boldsymbol\min}_{s\in S} C_s.\label{Optimization_Eq}\vspace{-0.05in}
\end{equation}
Here, $S$ denotes all the possible selection sets. We used an exhaustive search to address the optimization in \eqref{Optimization_Eq} because the search space is small. In large networks with a large design space, efficient search algorithms should be used to reduce the optimization complexity. Our proposed VL-selection algorithm is shown in Algorithm \ref{Algorithm_sel}. The algorithm is done for different VL-fault scenarios at design-time and the selection sets ($s^*$) are saved in routers for run-time use. For the baseline system (Fig.~\ref{Bseline}), where each chiplet has 4 VLs, there are 14 combinations of faults ($\binom{4}{1}+\binom{4}{2}+\binom{4}{3}$). Therefore, 14 VL addresses are saved in each router.\vspace{-0.05in}

\begin{algorithm}
\caption{Vertical-Link (VL) Selection in \textit{DeFT}}\label{Algorithm_sel}
\begin{algorithmic} 
\small
%\FOR {all possible fault scenarios (14 scenarios here)}
\FOR {each $s$ in all the possible selection sets ($s \in S$)}
\FOR{$v \in {VL}_C$}
\STATE $L_v \gets $ find load cost of $v$ (using \eqref{LoadCost_Eq})
\STATE $D_v \gets $ find distance cost of $v$ (using \eqref{DisCost_Eq})
\ENDFOR
\STATE $C_s \gets $ find the overall cost of selection set $s$ (using \eqref{OverallCost_Eq})
\IF {$C_s < C_s^*$}
\STATE $C_s^* \gets C_s$
\STATE Update the saved selection sets ($s^* \gets s$)
\ENDIF
\ENDFOR
%\ENDFOR
\end{algorithmic}
\end{algorithm}\vspace{-0.15in}

% •	Share traffic information in network
% •	Selection update procedure

\section{Evaluation and Simulation Results}\label{results}
\begin{figure*}[th]
  \centering
\includegraphics[width=1\textwidth]{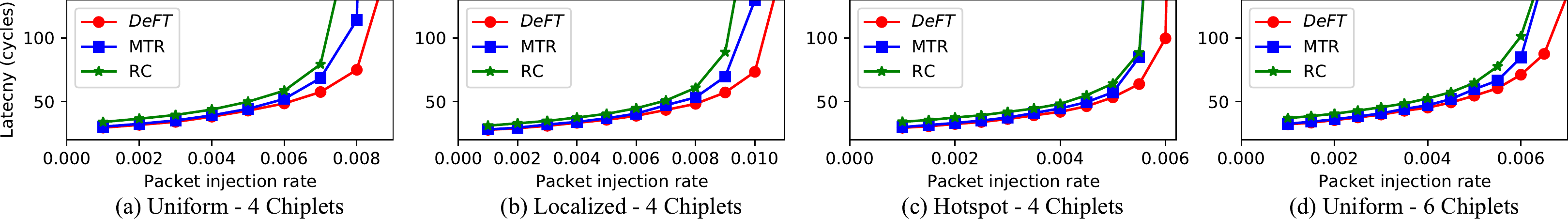}\label{uniform-6Chips}\vspace{-0.1in}
  \caption{\small Average latency comparison among \textit{DeFT}, MTR, and RC routing algorithms when applied to the network shown in Fig. 1 and under (a) Uniform, (b) Localized, and (c) Hotspot synthetic traffic patterns. (d) shows the same but for 6 chiplets and Uniform traffic.}\label{synthetic_latency}
 \vspace{-0.2in}
\end{figure*}
\begin{figure}[t]
\centering
\includegraphics[width=.42\textwidth]{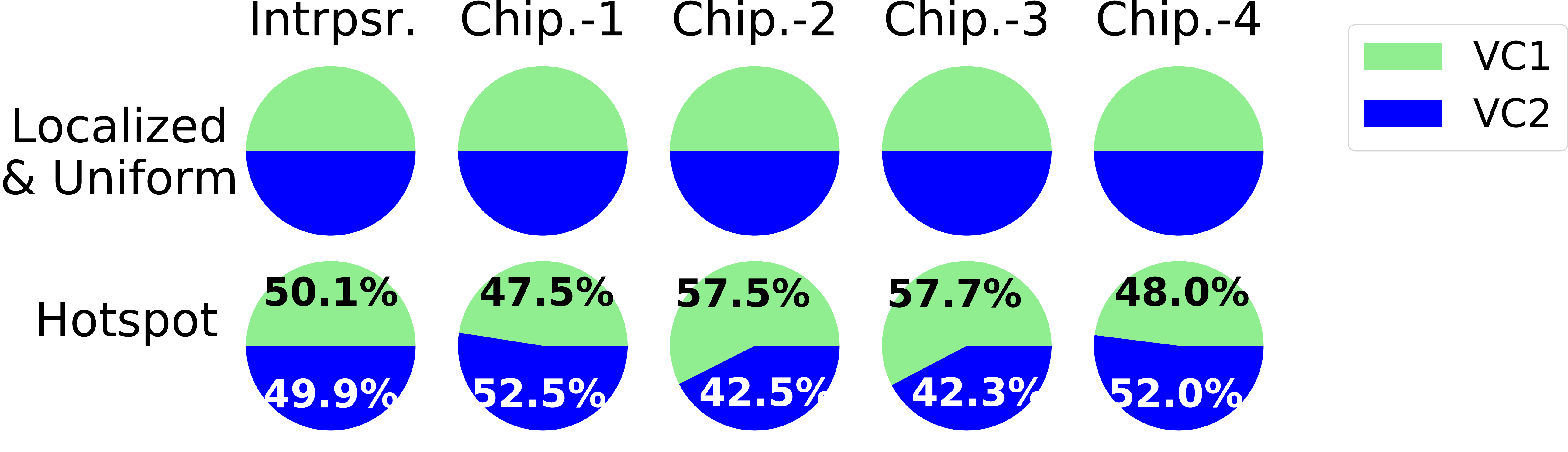}
\vspace{-0.1in}
\caption{\small VC utilization in \textit{DeFT} under synthetic traffic.}
\label{VCUtlization}
\vspace{-0.1in}
\end{figure}

\begin{figure}[t]
\centering
\includegraphics[width=.48\textwidth]{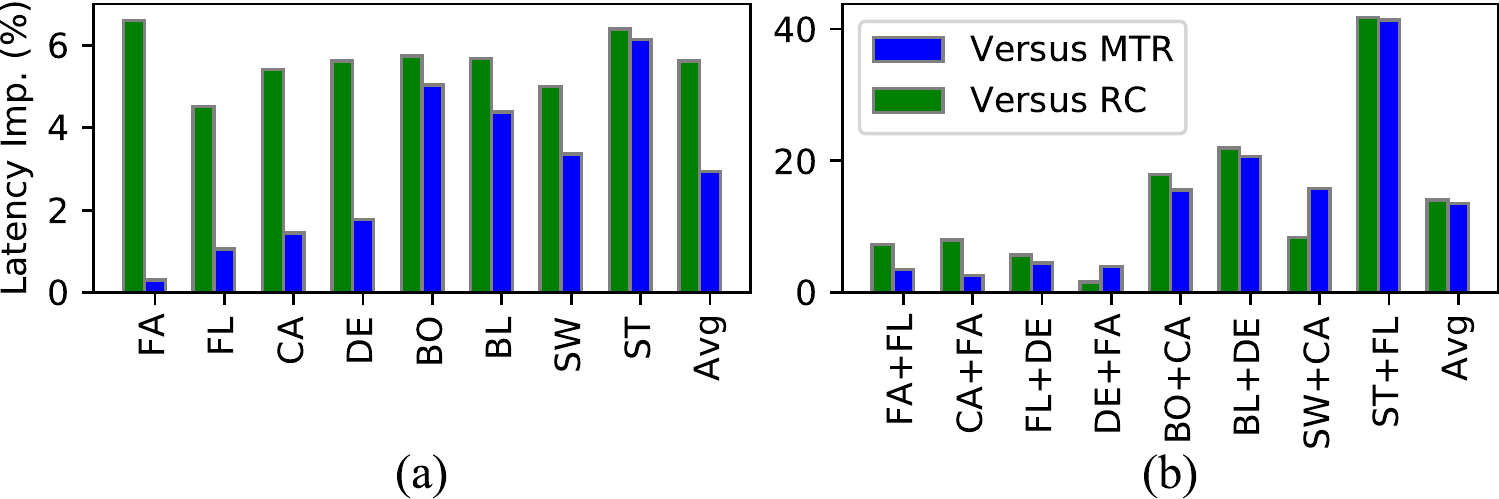}
\vspace{-0.1in}
\caption{\small Latency improvement under real-application traffic using (a) a single application, and (b) two applications simultaneously.}
\label{RealApp_latency}
\vspace{-0.25in}
\end{figure}

\begin{figure}[t]
\centering
\includegraphics[width=.48\textwidth]{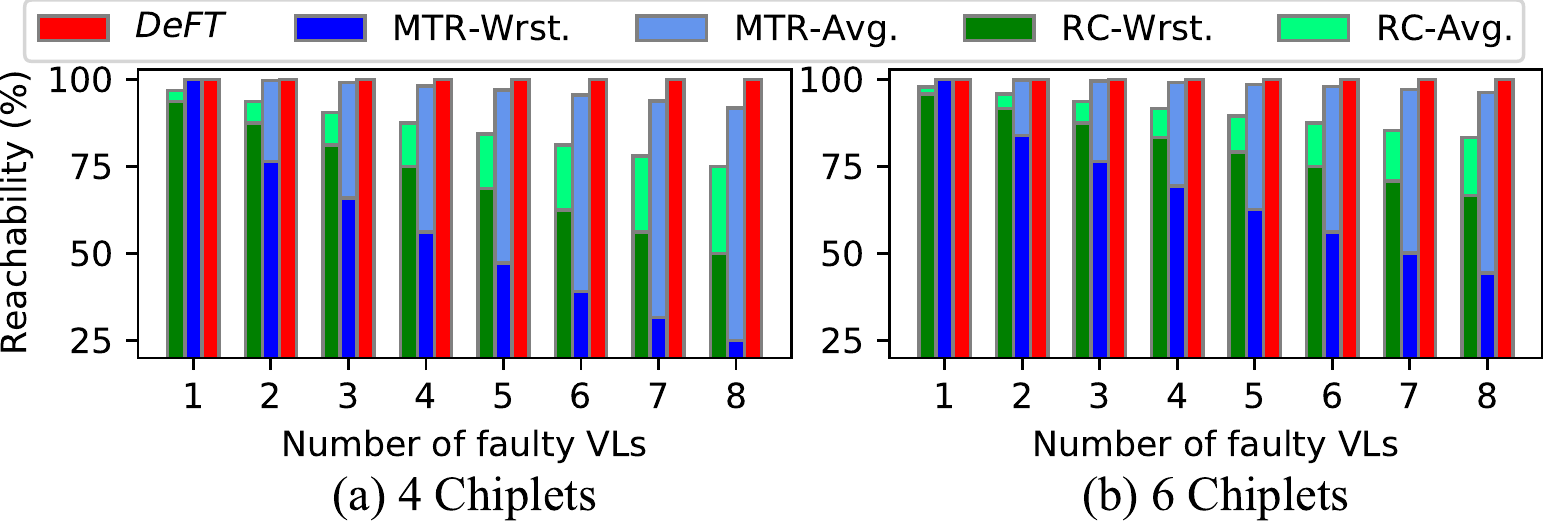}
\vspace{-0.1in}
\caption{\small Reachability in the presence of VL faults in a system with (a) 4 chiplets (total VLs$=$32), and (b) 6 chiplets (total VLs$=$48). Note that \textit{DeFT}-Wrst. and \textit{DeFT}-Avg. are the same (both shown by \textit{DeFT}).}
\label{Reachability}
\vspace{-0.25in}
\end{figure}

\subsection{Simulation Environment and Configuration}
We enhanced the Noxim simulator \cite{catania2016cycle}, which is a cycle accurate simulator for networks-on-chip, to support 2.5D chiplet systems as shown in Fig. \ref{Bseline}, where each chiplet is connected using four bidirectional VLs to the interposer. Each chiplet is connected to interposer with 4 VLs rather than fully connected, because reducing number of VLs alleviates fabrication costs. To further reducing VLs and improving fabrication cost serialization can be employed \cite{pasricha2009exploring}. According to \cite{yin2018modular}, for a 4$\times$4 chiplet, placing the four VLs on the borders of the chiplet is an optimal choice when considering hardware complexity and network latency. Note that \textit{DeFT} efficiency is independent of the placement and density of VLs as it does not use turn models for achieving inter-chiplet deadlock-freedom. Besides the baseline 4-chiplet system, we simulated a system with 6 chiplets to show how \textit{DeFT}'s efficiency scales with system size. We compare \textit{DeFT} against state-of-the-art routing algorithms for chiplet-based systems, MTR \cite{yin2018modular} and RC \cite{majumder2020remote}. While MTR and RC have no requirement for the number of VCs, two VCs are used in all the algorithms to have a fair comparison with \textit{DeFT} (i.e., considering a single VC will make MTR and RC perform worse). A packet size of eight flits and a buffer size of four flits are considered, where a flit width is 32 bits. We performed offline VL-selection optimization considering Uniform traffic, the most pessimistic assumption, while our simulations include different traffic scenarios. Including traffic information in the offline optimization results in further improvements. \vspace{-0.05in}

\subsection{Latency Analysis}
% •	Compare latency with 2.5D routing algorithms in fault free scenarios
% •	One app and 2 app latency results in fault free network
The latency analysis for synthetic traffic is shown in Fig.~\ref{synthetic_latency}. Compared to MTR and RC, \textit{DeFT} has the lowest average latency because of its balanced VL selection and VC utilization. In Localized traffic, shown in Fig.~\ref{synthetic_latency}(b), for 40\% of the packets, the source and destination are on the same chiplet, i.e., intra-chiplet packets. \textit{DeFT} shows a low latency because the VC utilization for intra-chiplet packets is highly distributed (see Theorem.\ref{theorem1}). In Hotspot traffic, shown in Fig.~\ref{synthetic_latency}(c), \textit{DeFT} shows slightly lower improvement because incoming packets to chiplets are allowed to only use the second VC, which can result in a brief back-pressure. However, we used a relatively high rate of hotspot in our simulation (3 hotspot points with 10\% rate on each). As shown in Fig.~\ref{synthetic_latency}(d), where the 6-chiplet system is considered, \textit{DeFT} performance is not limited by system size. The distribution of VC utilization for synthetic traffic patterns are shown in Fig.~\ref{VCUtlization}. VC utilization is balanced (50\% with less than 0.4\% tolerance) in Uniform and Localized traffic, hence results for both are presented in the same chart. In Hotspot traffic, although hotpot rate is relatively high, deviation in VC utilization is less than 8\%. Balanced VC utilization in \textit{DeFT} offers a low average latency, while avoiding deadlock.

To consider real-application traffic, we generated traffic from PARSEC benchmarks \cite{bienia2011benchmarking} using GEM5 \cite{binkert2011gem5}, and simulated the generated traffic with our chiplet-based Noxim simulator. We simulated the benchmarks in full-system mode with 64 x86 cores, four coherence directories, and four shared L2 cache banks (each core also has a private L1 cache). The latency improvement in eight PARSEC applications (blackscholes, bodytrack, canneal, dedup, facesim, fluidanimate, streamcluster, and swaptions) is shown in Fig.~\ref{RealApp_latency}, where the first two letters of each application are used on the x-axis. Because of low congestion in the network when running a single application, we also simulated two applications running simultaneously, where each application executed on 32 cores (see Fig.~\ref{RealApp_latency}(b)). On average, \textit{DeFT} shows more improvement when multiple applications are considered  due to the higher chance of network congestion. On the x-axis in Fig.~\ref{RealApp_latency}(b), the two-application combinations are sorted based on traffic load, from low (FA+FL) to high (ST+FL). For high traffic loads, \textit{DeFT} shows a notable improvement of up to 40\% compared to both MTR and RC. \vspace{-0.05in}

% •	Compare reachability with 2.5D routing algorithms in vertical link faults
% •	Compare latency in fault scenarios (sensitivity analysis)
\subsection{Fault-Tolerance Analysis}
To assess \textit{DeFT}'s ability to tolerate faults, we analyze network reachability in the presence of faults, similar to \cite{salamat2016resilient}. In VL-fault scenarios, reachability is defined as the ratio of packets that can be successfully routed, to the total number of injected packets. As shown in Fig.~\ref{Reachability}(a), \textit{DeFT} achieves complete (i.e., 100\%) reachability for the considered fault-injection rates. We injected all combinations of fault patterns excluding those that disconnected chiplets completely (i.e., when all VLs of a chiplet are faulty). The figure shows both the average and the worst-case reachability under an injected fault rate of 3.125\% to 25\% (1--8 faulty VLs). Considering the average case, \textit{DeFT} improves the network reachability by 3.1--25\% compared to RC, and by 0--8.1\% compared to MTR. In the worst case, \textit{DeFT} improves the network reachability by 6.25--50\% compared to RC, and by 0--75\% compared to MTR. Fig.~\ref{Reachability}(b) shows network reachability with the 6-chiplet system. Here, \textit{DeFT} can achieve 100\% reachability for the fault injection rates shown, while MTR provide 100\% reachability for only the low fault-rate scenario with 2.1\% faults (one faulty VL). Also, RC cannot tolerate any faults. The restricted turns in MTR and the permission network of RC limit their VL selection and, therefore, fault-tolerance.    
% Numbers are based on worst case
%Although in chiplet systems, DeFT addresses fault tolerant VL selection for the first time, fault-tolerant VL selections for 3D NoCs are already presented \cite{salamat2016resilient,coelho2019fl,niazmand2016logic} in which closest VL is assumed as a replacement for faulty VLs. 

\begin{figure}[t]
\centering
\includegraphics[width=.47\textwidth]{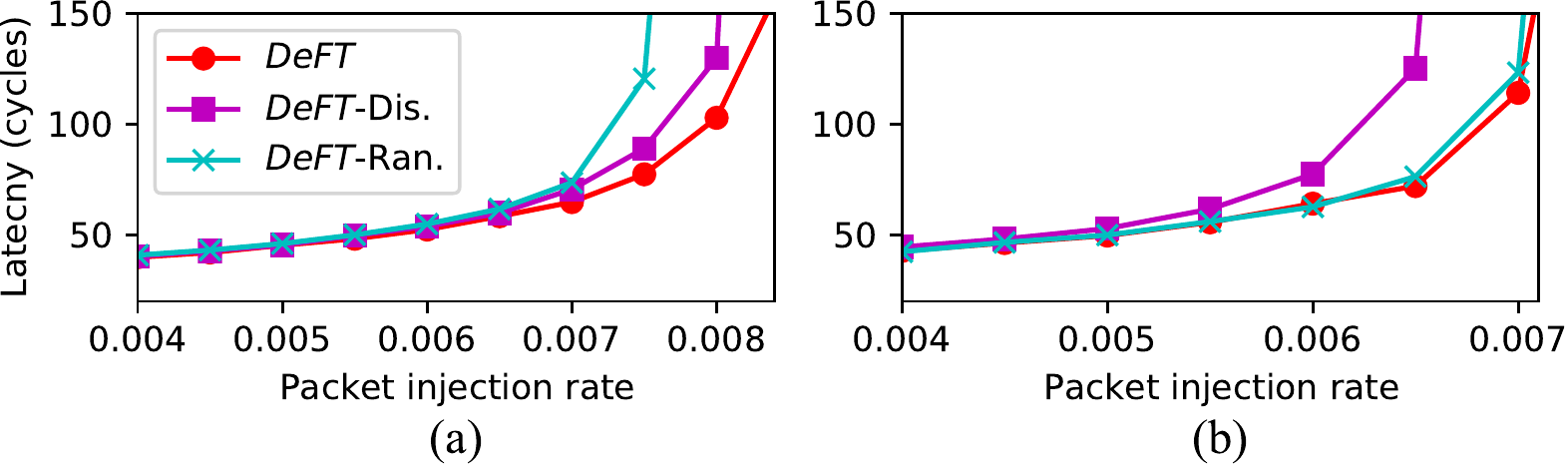}\hspace{-0.2in}
\vspace{-0.1in}
\caption{\small Average latency in \textit{DeFT} with different VL-selection strategies and fault-injection rates for a 4-chiplet system. (a) VL fault-rate of 12.5\% (4 faulty VLs), and (b) VL fault-rate of 25\% (8 faulty VLs).}
\label{latencyVLFaults}
\vspace{-0.15in}
\end{figure}
\begin{table}[t!]
\caption{Area and power analysis of \textit{DeFT}, MTR, and RC}
\vspace{-0.1in}
\centering
\small
\begin{tabular}{|c||c|c|c|c|}
\hline
   & MTR & RC\textsubscript{non-bndry} & RC\textsubscript{bndry} & \textit{DeFT}\\
\hline
\hline
%Router $(\mu m^2)$ & 45687 & 46472 & 51793 & 46460\\
Router area ($\mu$m$^2$) & 45878 & 46663 & 51984 & 46651\\
\hline
%CDA$^*$  & $37619$ & \multicolumn{1}{c|}{$5.8\%$} \\
Norm. router area & 1  & 1.017 & 1.133 & 1.016 \\
\hline
Router power (mW) &  11.644 & 11.76  & 12.841  & 11.693 \\
\hline
%CDA$^*$  & $37619$ & \multicolumn{1}{c|}{$5.8\%$} \\
Norm. router power & 1  & 1.009 & 1.102 & 1.004 \\
\hline
\end{tabular}
\label{hardware}
\vspace{-0.25in}
\end{table}

Fig.~\ref{latencyVLFaults} shows the impact of VL faults on latency. Here, MTR and RC are not considered because they are unable to offer a complete reachability under fault scenarios. Moreover, in addition to using \textit{DeFT} with the proposed VL selection (\textit{DeFT} in Fig.~\ref{latencyVLFaults}),  we include the average latency in \textit{DeFT} with distance-based (\textit{DeFT}-Dis.)---a common approach in 3D networks \cite{salamat2016resilient}---and random (\textit{DeFT}-Ran.)---a VL is selected randomly among eligible VLs---VL-selection strategies. Here, \textit{DeFT} with distance-based selection (\textit{DeFT}-Dis. in Fig.~\ref{latencyVLFaults}) means that the routing is \textit{DeFT} while VL selection is based on distance instead of our proposed selection. \textit{DeFT} shows significantly lower latency than those with random- and distance-based VL selections under 12.5\% and 25\%  fault injection rates, respectively. Random selection offers a good load distribution when the number of faulty VLs is large (e.g., under 25\% fault rates). But random selection imposes latency overhead under 12.5\% fault rates (see Fig.~\ref{latencyVLFaults}(a)). As shown in Fig.~\ref{latencyVLFaults}(b), latency in \textit{DeFT} is still lower under 25\% fault rates. \vspace{-0.05in}\par

% •	Compare reliability with 2D algorithms in horizontal link faults
\subsection{Hardware and Power Analysis}
\label{hardwareSec}
% •	Analysis hardware complexity of routing
% •	Hardware complexity of selection
We used Cadence Genus and ORION 3.0 \cite{kahng2015orion3} to estimate router area and power at the 45~nm technology and considering a 1~GHz clock frequency. Table~\ref{hardware} compares the area and power estimation of a six-port \textit{DeFT} router with routers used in MTR and RC. Note that the hardware implementations of the non-boundary and boundary routers are different in RC, hence we report both separately in the table. From Table.~\ref{hardware}, \textit{DeFT} imposes less than 2\% and 1\% hardware and power overhead, relatively, at most compared to related work. The small overhead includes the logic for VN-assignment algorithm and look-up tables to store data for fault-tolerant VL selection.

%Fig.~\ref{energy}(a) shows energy analysis normalized to MTR for 100K packets injected under Uniform traffic and 0.01 packets/node/cycle injection rate. \textit{DeFT} achieves lower energy by 6.1\% and 10.7\% compared to, respectively, MTR and RC as it can route the injected packets faster. In addition, RC uses extra buffers for deadlock-freedom, which increases energy costs. MTR sends packets to non-minimal VLs to guarantee deadlock-freedom, while \textit{DeFT} is free to select any VLs. Also, under 12.5\% and 25\% VL-fault scenarios, shown in Figs.~\ref{energy}(b) and~\ref{energy}(c), compared to random (\textit{DeFT}-Ran.) and distance-based (\textit{DeFT}-Dis.) VL selection, \textit{DeFT} with its congestion-aware VL selection is able to achieve lower energy.
% \begin{figure}[t]
% \centering
% \includegraphics[width=.47\textwidth]{Energy.pdf}
% \vspace{-0.1in}
% \caption{\small Normalized energy comparison for a 4-chiplet network with (a) no faults, (b) 12.5\% faulty VLs, and (c) 25\% faulty VLs.}
% \label{energy}
% \vspace{-0.25in}
% \end{figure}

\vspace{-0.05in}

\section{Conclusion}
This paper presents the first deadlock-free and fault-tolerant routing algorithm, called \textit{DeFT}, for 2.5D chiplet networks. \textit{DeFT} offers VC-based deadlock-freedom where not only are packets free to select any VL, but also VC utilization is efficiently balanced. Freedom in VL-selection allows \textit{DeFT} to tolerate any pattern of VL faults. To improve network congestion in VL-fault scenarios, \textit{DeFT} employs a dynamic traffic-aware VL selection to enhance run-time routing efficiency. Compared to the state-of-the-art routing algorithms, our simulation results show that \textit{DeFT} improves network reachability by up to 75\% with a fault rate of up to 25\% and reduces the network latency by up to 40\% for multi-application execution scenarios with less than 2\% area overhead. These results highlight the promise of \textit{DeFT} for improving emerging chiplet systems. %While this paper addresses, for the first time, the critical problem of VL faults in 2.5D chiplet systems, the addition of horizontal-link faults will be considered in our future work.
\vspace{-0.05in}

\section*{Acknowledgment}
This work was supported by the National Science Foundation (NSF) under grant number CNS-2046226.

\vspace{-0.05in}

\bibliographystyle{IEEEtran}
\bibliography{ref}

\end{document}